\documentclass[twocolumn,aps,pra,superscriptaddress,nofootinbib]{revtex4-2}

\usepackage{pifont}

\usepackage{color, colortbl}
\definecolor{Gray}{gray}{0.2}

\usepackage{graphicx}
\usepackage[dvipsnames,svgnames]{xcolor}
\usepackage{bm,bbm}
\usepackage{braket}
\usepackage{amsthm,amsmath,amssymb}
\usepackage{enumerate}
\usepackage{booktabs,multirow}
\usepackage{mathtools,bbding}
\usepackage[percent]{overpic}
\usepackage{microtype}
\usepackage{array,adjustbox,afterpage}
\usepackage[T1]{fontenc}
\usepackage[utf8]{inputenc}
\usepackage{tikz}
\usetikzlibrary{calc}
\usepackage[english]{babel}
\usepackage{newtxtext,newtxmath}

\usepackage{lipsum}
\usepackage{pifont}

\usepackage[linesnumbered,ruled]{algorithm2e}  %delete ruled and write boxed for box

\makeatletter
\renewcommand{\fnum@figure}{FIG. \thefigure} % Figure --> FIG.
\makeatother

\usepackage[bookmarks=false, colorlinks=true, linkcolor=blue, citecolor=purple, urlcolor=purple]{hyperref}
\usepackage{cleveref}
\usepackage[normalem]{ulem}

\usepackage{lipsum}
\usepackage{makecell}

\usepackage{caption}
\usepackage{subcaption}
\usepackage{ragged2e}

\Crefname{subfigures}{figure}{figures}
\Crefname{subfigures}{Figure}{Figures}

\newcommand{\xdownarrow}[1]{{\left\downarrow\vbox to #1{}\right.\kern-\nulldelimiterspace}} % for long downarrow ..

\hypersetup{
    breaklinks=true,   % splits links across lines
    colorlinks=true,   % displays links as colored text instead of blocks
    pdfusetitle=true,  % \title and \author values into pdf metadata etc.
}

\begin{document}

\title{The physical basis of information flow in neural matter:\\ a thermocoherent perspective on cognitive dynamics}
\author{Onur Pusuluk}
\email{onur.pusuluk@gmail.com}
\affiliation{Faculty of Engineering and Natural Sciences, Kadir Has University, 34083, Fatih, Istanbul, T\"{u}rkiye}
\date{April 2026}

\begin{abstract}

Information flow is central to contemporary accounts of cognition, yet its physical basis in living neural matter remains poorly specified. Here, we develop a multiscale resource-theoretical framework motivated by the \textit{thermocoherent effect}, where heat flow is reciprocally coupled to a delocalized information flow carried by shared coherence and not reducible to local subsystem variables. Extending this line of work in light of recent results on correlation-enabled Mpemba-type thermal relaxation, we argue that the operational relevance of correlations depends less on their taxonomy than on their dynamical accessibility under the underlying interaction geometry. Relational structure encoded in the state of a single composite system -- including quantum entanglement, quantum discord, and classical correlations -- may therefore act as a usable physical resource that remains hidden from local subsystem descriptions. We propose that electrical, chemical, ionic, and thermal transport processes in neural matter may, under suitable microscopic conditions, generate or transduce partially hidden relational resources whose mutual coupling can progressively build larger-scale thermocoherent organization across spatial or spatiotemporal partitions in neural tissue. Ion-channel interfaces, hydrogen-bonded proton networks, aromatic $\pi$-electron architectures, and phosphate-rich motifs emerge as plausible substrate classes in which such resources may arise, become transiently accessible under environmental coupling, and leave coarse-grained signatures in neural dynamics. In this perspective, electromagnetic field-based binding proposals are reinterpreted not as primary microscopic carriers of cognition, but as emergent mesoscale coordination layers shaped by deeper transport-coupled thermodynamic constraints. The resulting picture is neither a claim of macroscopic quantum cognition nor a reduction of cognition to abstract coding, but a falsifiable framework in which microscopic relational resources can bias transport, relaxation, signaling, and cross-scale neural coordination.

\end{abstract}

\maketitle

\section{Introduction}

\begin{figure*}[t]
    \centering
    \includegraphics[width=.6\textwidth]{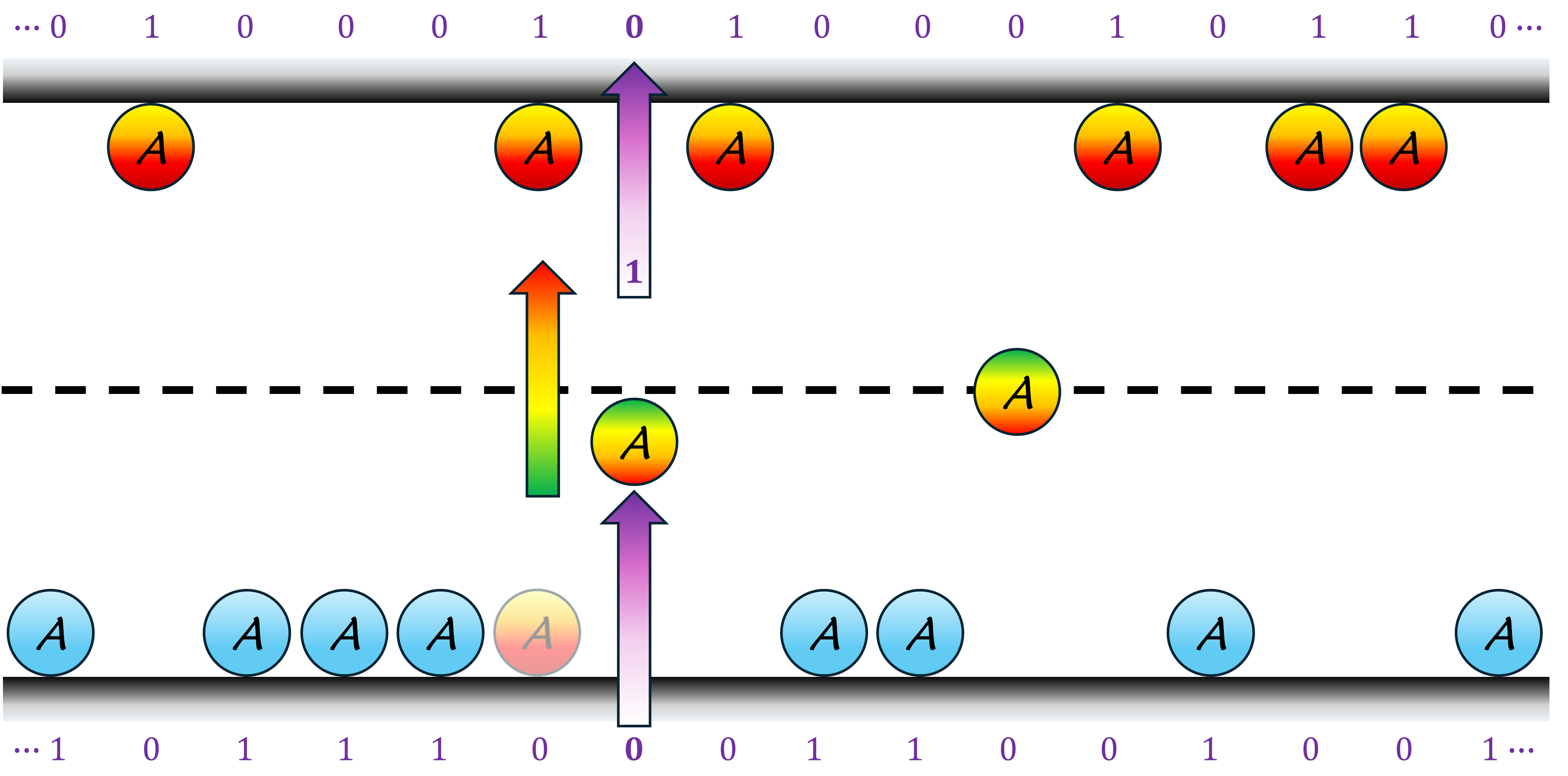}
    \caption{\justifying A schematic illustration of a carrier-based local picture of information flow. In this conventional view, the flow of information is identified with the motion or transduction of locally encoded physical carriers (here represented by binary occupancy and color-coded local states). This intuition is often useful at coarse-grained levels, but it does not exhaust the physically relevant possibilities in nonequilibrium composite systems, where information can instead be stored or transferred in relational structure not reducible to any single local current.}
    \label{Fig::LocalInfoFlow}
\end{figure*}

Information flow plays a central role in contemporary accounts of cognition, yet what kind of physical entity or process it corresponds to is still unclear. In many descriptions, it is treated either as a largely abstract, phenomenological, or coding-level notion~\cite{Friston2010, 2012_PhysOfLifeRew_InfoFlow}, or else as something fully specified by the nonequilibrium organization of the material substrates that support it~\cite{2015_FrontInPscho_InfoThermo2Neuro, 2024_Entropy_InfoThermo2Neuro} (see Fig.~\ref{Fig::LocalInfoFlow}). Taken in isolation, these viewpoints can lead to complementary limitations: either detaching information from the physical transport processes that implement and constrain it, or identifying it too directly with the local physical currents that accompany it. Recent developments in quantum information thermodynamics~\cite{gour2015resource,goold2016role,lostaglio2019introductory,torun2023compendious} suggest a more nuanced possibility: information is a \emph{physical resource}~\cite{coecke2016mathematical,chitambar2019quantum} that need not be reducible to local subsystem properties and can, under suitable conditions, be distributed or shared through relational structure, including correlations. This in turn opens the possibility of grounding information flow physically without reducing it to subsystem-local transport alone. It also brings into sharper focus a reciprocal question that remains conceptually underdeveloped in many cognitive and neuroscientific accounts: once physically instantiated, how information-bearing organization can in turn constrain, redirect, or modulate subsequent material dynamics.

A particularly clear example is provided by the thermocoherent effect~\cite{Pusuluk_2021_PRR}. In that setting, heat flow is reciprocally coupled to a \emph{delocalized information flow} carried by quantum coherence~\cite{baumgratz2014quantifying, streltsov2017colloquium} shared between degenerate energy levels, so that the latter is not localized within either subsystem even though it can directly influence energy transport. This feature distinguishes the thermocoherent effect from more generic coherence-assisted thermodynamic phenomena~\cite{AHF_2008_PRE_Partovi, Lutz2009, AHF_2010_PRE_JenningsAndRudolph, 2016_Entropy_Ozgur, 2018_PRE_RoleOfQCohInHT, 2019_npj, 2019_PRE_Ozgur_Multiatom, 2019_PRL_Esposito_TDofInfoFlow, 2019_PRA_CohNeeded4HF, AHF_2019_NatCommun_Lutz, 2019_OSID, AHF_2019_PhysRevResearch_Petruccione, 2019_PRL_Esposito_PiWithBathCoh, AHF_2020_arXiv_IonTraps}: its key novelty is not merely that coherence modifies heat flow, but that an explicitly delocalized information flow emerges alongside it, remains reciprocally coupled to the heat current, and thereby mediates the redistribution of relational structure across subsystems, opening a distinct route by which hidden organization can feed back onto energy transport. In this sense, the thermocoherent effect may be understood as a prototype of a broader class of transport-coupled hidden relational effects. Here, the thermocoherent label is not restricted to shared coherence alone.

More recent resource-theoretic analyses, including correlation-enabled Mpemba-type relaxation scenarios~\cite{Pusuluk_qMpemba}, broaden this lesson further. In such cases, thermodynamically active hidden structure need not be restricted to coherence-based resources alone: under suitable interaction geometry and spectral accessibility, operationally relevant structure may also be carried by \emph{classical correlations}~\cite{Vedral-2001, modi2010unified, modi2012classical} encoded in the state of a single composite system, even when the local marginals remain thermodynamically indistinguishable. Once such hidden relational structure becomes dynamically accessible, it can alter relaxation ordering and effective cooling behavior.

This claim requires a precise use of the term ``classical.'' Here, ``classical'' does not refer to ensemble-level statistical dependence, trial-averaged covariance, or a merely epistemic lack of knowledge. Rather, it denotes physically instantiated relational structure in an individual composite system, as represented by its density operator. Nonclassical correlations~\cite{modi2012classical, adesso2016measures}, including quantum entanglement~\cite{1998_ContempPhys_Ent, Plenio2007, 2008_RevModPhys_ManyBodyEnt} and quantum discord~\cite{Zurek-2002, modi2010unified, bera2017quantum}, are understood in the same state-level sense. The present framework therefore concerns relational resources internal to single composite systems, not coarse-grained statistical correlations across repeated trials. This distinction matters operationally, because systems with indistinguishable local marginals may still differ substantially in hidden relational structure that becomes dynamically relevant only under suitable couplings.

The scope of this perspective extends beyond single-time state-level descriptions. It need not be confined to relational structure across spatial partitions at a single instant. Depending on the operational setting, relevant hidden structure may also appear in temporally ordered or multi-time relational form~\cite{2015_SRep_VV, 2018_PRL_ProcessTensor} that is inaccessible even to instantaneous joint-system descriptions yet still constrains later transport, response, or route selection. In history-dependent settings, such process-level relational structure may also provide a natural language for discussing memory-like behavior and certain forms of non-Markovianity~\cite{2023_Frontiers_Review}.

Against this background, the aim of the present work is to extend this substrate-agnostic thermodynamic perspective toward biological and neural matter without introducing a new neuroscientific primitive or presupposing macroscopic quantum cognition. We ask whether electrical~\cite{HodgkinHuxley1952MembraneCurrent, Bean2007}, chemical~\cite{Destexhe1994, Sdhof2013, Wang2021}, ionic~\cite{Hille2001}, and thermal~\cite{Attwell2001, Harris2012, Kiyatkin2019} transport processes in living systems can support forms of relational organization whose redistribution, dynamical accessibility, or temporal persistence become operationally relevant in ways analogous to the thermocoherent and correlation-enabled effects discussed above. Our primary focus here is neural matter, not because the underlying framework is intrinsically neuron-specific, but because neural tissue provides an especially rich and experimentally consequential setting in which multiple transport channels coexist, interact across scales, and can plausibly render hidden relational structure dynamically accessible.

\section{State-level framework and preliminaries} \label{sec:framework}

Before turning to operational archetypes and candidate substrate classes, we first introduce the minimal formal distinctions used throughout the rest of this work. At this stage, the aim is simply to identify how relational structure can be encoded in composite descriptions, how it may remain hidden from subsystem-reduced views, and under what conditions it can become operationally accessible.

\subsection{Relational states and hidden structure}

Consider two subsystems \(A\) and \(B\) with basis states \(\{|0\rangle_A,|1\rangle_A\}\) and \(\{|0\rangle_B,|1\rangle_B\}\). We take their local states to be diagonal in these bases,
\begin{equation}
\begin{aligned}
\hat{\rho}_A &= p\,|0\rangle_A\langle 0| + (1-p)\,|1\rangle_A\langle 1|, \\
\hat{\rho}_B &= q\,|0\rangle_B\langle 0| + (1-q)\,|1\rangle_B\langle 1|,
\end{aligned}
\end{equation}
so that \(p\) and \(q\) are the populations of the local \(|0\rangle\) states, while \(1-p\) and \(1-q\) are the populations of the local \(|1\rangle\) states.

For pedagogical concreteness, \(A\) and \(B\) may be interpreted as two localized physical modes relevant to a given substrate class, such as proton-occupancy sites along a hydrogen-bond network or ion-occupancy sites within a channel selectivity filter. In realistic neural settings, the local basis states \(|0\rangle\) and \(|1\rangle\) may represent absence/presence, unoccupied/occupied, or spin-down/spin-up configurations depending on the context. The formal structure developed below is independent of the particular interpretation.

The product state \(\hat{\rho}^{\mathrm{uc}}_{AB} = \hat{\rho}_A \otimes \hat{\rho}_B\) is called uncorrelated and reads
\begin{equation}
\hat{\rho}^{\mathrm{uc}}_{AB} =
\begin{pmatrix}
pq & 0 & 0 & 0\\
0 & p(1-q) & 0 & 0\\
0 & 0 & (1-p)q & 0\\
0 & 0 & 0 & (1-p)(1-q)
\end{pmatrix}
\end{equation}
in the computational product basis \(\{|00\rangle,|01\rangle,|10\rangle,|11\rangle\}\). This state contains no relational structure beyond the local marginals.

\paragraph{Classically correlated states.}

A simple classically correlated state with the same local marginals can then be written as
\begin{equation}
\hat{\rho}^{\mathrm{cc}}_{AB}({\chi}) = \hat{\rho}^{\mathrm{uc}}_{AB} + \Delta_{\chi},
\end{equation}
with
\begin{equation}
\Delta_{\chi}=
\begin{pmatrix}
\chi & 0 & 0 & 0\\
0 & -\chi & 0 & 0\\
0 & 0 & -\chi & 0\\
0 & 0 & 0 & \chi
\end{pmatrix},
\end{equation}
where \(\chi\) is restricted by positivity. The reduced states of \(\hat{\rho}^{\mathrm{cc}}_{AB}({\chi})\) are still \(\hat{\rho}_A\) and \(\hat{\rho}_B\), but the joint probabilities are no longer products of the marginals. Thus, the relational structure of the composite system has changed even though the local descriptions have not. For every nonzero allowed value of \(\chi\), the state \(\hat{\rho}^{\mathrm{cc}}_{AB}({\chi})\) is non-product and carries genuine classical correlation~\cite{Vedral-2001, modi2010unified, modi2012classical}.

This becomes especially transparent when \(p=q\) and \(\chi=\chi_{\max}=p(1-p)\). In that case,
\begin{equation}
\hat{\rho}^{\mathrm{cc}}_{AB}({\chi_{\max}})=
\begin{pmatrix}
p & 0 & 0 & 0\\
0 & 0 & 0 & 0\\
0 & 0 & 0 & 0\\
0 & 0 & 0 & 1-p
\end{pmatrix},
\end{equation}
which is supported only on \(|00\rangle\) and \(|11\rangle\). Two systems prepared in states \(\hat{\rho}^{\mathrm{uc}}_{AB}\) and \(\hat{\rho}^{\mathrm{cc}}_{AB}(\chi_{\max})\) may look identical locally and yet differ maximally in their joint relational organization.

\paragraph{Quantum discordant states.}

A second example shows how nonclassical relational structure can be introduced without modifying the local marginals. For \(p=q\), consider
\begin{equation}
\hat{\rho}^{\mathrm{qd}}_{AB}({\lambda})=\hat{\rho}^{\mathrm{uc}}_{AB}+\Delta_{\lambda},
\end{equation}
where
\begin{equation}
\Delta_{\lambda}=
\begin{pmatrix}
0 & 0 & 0 & 0\\
0 & 0 & \lambda & 0\\
0 & \lambda^* & 0 & 0\\
0 & 0 & 0 & 0
\end{pmatrix},
\end{equation}
with \(\lambda\) again restricted by positivity. The off-diagonal block couples \(|01\rangle\) and \(|10\rangle\), introducing shared coherence in the single-excitation sector while leaving the local populations unchanged. For every nonzero allowed value of \(\lambda\), this state carries nonzero shared coherence and, in this construction, nonzero quantum discord~\cite{Zurek-2002, modi2010unified, bera2017quantum}, making it a minimal representative of a nonclassical but non-entangled relational resource.

\paragraph{Quantum entangled states.}

An entangled counterpart may then be constructed by adding shared coherence to the classically correlated background,
\begin{equation}
\hat{\rho}^{\mathrm{qe}}_{AB}({\mu})=\hat{\rho}^{\mathrm{cc}}_{AB}({\chi_{\max}})+\Delta_{\mu},
\end{equation}
with
\begin{equation}
\Delta_{\mu}=
\begin{pmatrix}
0 & 0 & 0 & \mu\\
0 & 0 & 0 & 0\\
0 & 0 & 0 & 0\\
\mu^* & 0 & 0 & 0
\end{pmatrix},
\end{equation}
where \(\mu\) is restricted by positivity. In this case, the coherence resides in the \(|00\rangle \leftrightarrow |11\rangle\) sector. For every nonzero allowed value of \(\mu\), the resulting state is entangled~\cite{1998_ContempPhys_Ent, Plenio2007, 2008_RevModPhys_ManyBodyEnt} while preserving the same local marginals.

The important point is that \(\hat{\rho}_A\) and \(\hat{\rho}_B\) can be kept fixed while the global relational structure changes qualitatively across \(\hat{\rho}^{\mathrm{cc}}_{AB}\), \(\hat{\rho}^{\mathrm{qd}}_{AB}\), and \(\hat{\rho}^{\mathrm{qe}}_{AB}\). The three parameters \(\chi\), \(\lambda\), and \(\mu\) therefore play parallel pedagogical roles: each leaves the subsystem marginals unchanged while activating a distinct relational sector of the global state -- classical correlation, discordant separability, and entanglement, respectively.

These examples illustrate the minimal distinction used throughout the rest of this work. Two composite systems may share identical local marginals while differing in the relational structure encoded in their joint density operators. Whether such differences remain dynamically silent or become operationally relevant depends on the interaction geometry and spectral accessibility of the coupling.

\subsection{Delocalized information flow}
\begin{figure*}[t]
    \centering
    \includegraphics[width=.6\textwidth]{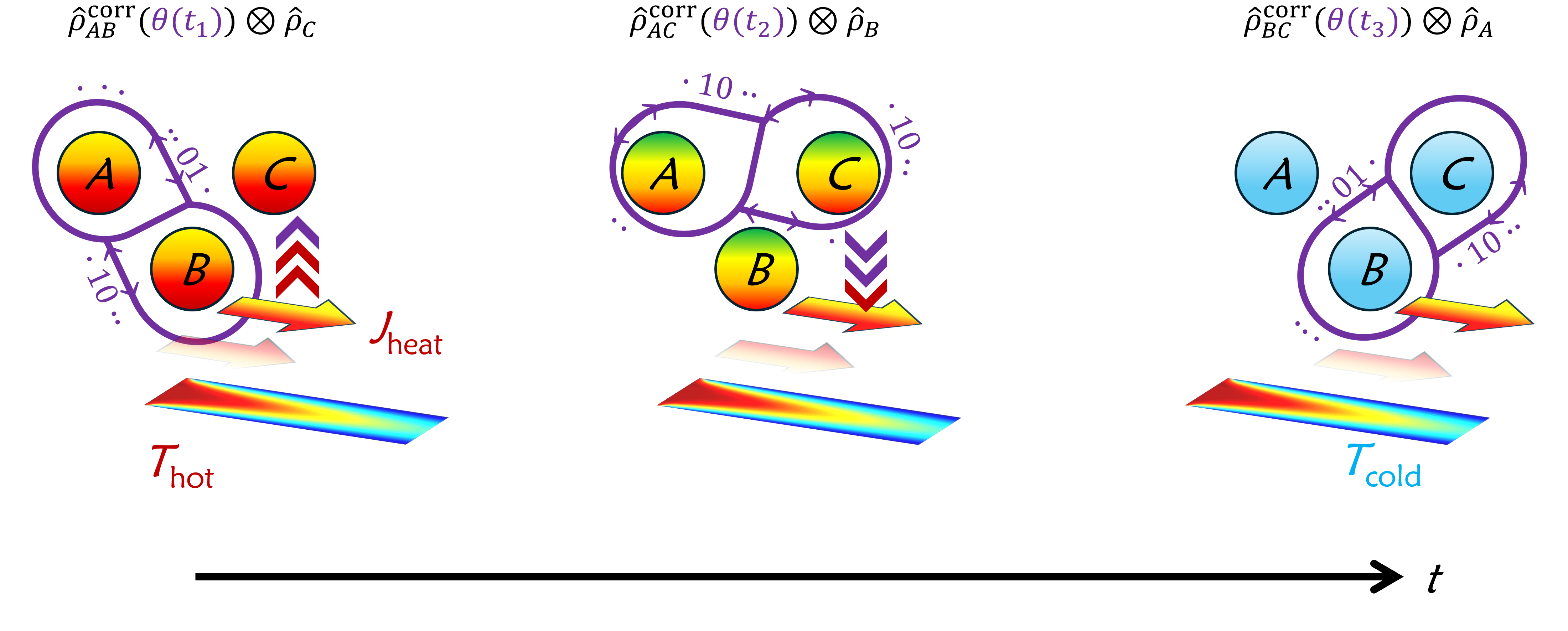}
    \caption{\justifying Schematic illustration of a delocalized information flow in a nonequilibrium three-subsystem setting. Relational support is redistributed across subsystem partitions during relaxation, alongside an accompanying heat current \(J_{\mathrm{heat}}\) and thermal gradient. The subsystem colors indicate local temperatures, while the superscript \(\mathrm{corr}=\{\mathrm{cc},\mathrm{qd},\mathrm{qe}\}\) denotes the pairwise relational structure. The parameter \(\theta\) takes the values \(\chi\), \(\lambda\), and \(\mu\), respectively.}
    \label{Fig::DelocalInfoFlow}
\end{figure*}

The examples above motivate a more general notion of information flow than one based solely on subsystem-local variables. When an interaction redistributes, transfers, reshapes, or converts accessible relational structure across different partitions of a larger composite state, what changes is not merely a local observable but the pattern of joint organization itself. In this sense, one may speak of a \emph{delocalized information flow}. In correlation-dominated settings, this may also appear as a redistribution of correlations across subsystem partitions, but the notion is broader than correlation flow alone.

The term \emph{delocalized} is important here. It does not simply mean that a signal is spatially spread out, nor does it imply robust long-lived macroscopic quantum coherence. Rather, it means that the relevant dynamical resource is carried by relational degrees of freedom of a composite state rather than by a subsystem-local variable. In the two-subsystem examples above, this resource is encoded in the correction terms \(\Delta_{\chi}\), \(\Delta_{\lambda}\), and \(\Delta_{\mu}\). In larger multipartite systems, one may caricature a delocalized information flow as the redistribution of relational structure across subsystem partitions, as illustrated schematically in Fig.~\ref{Fig::DelocalInfoFlow}: relational support initially concentrated in the \(AB\) sector may, under the dynamics, be partially redistributed into the \(AC\) sector, the \(BC\) sector, or into overlapping higher-order sectors. The flow in this picture is not the motion of a local quantity through space, but the migration, reshaping, or transfer of operationally accessible relational support across the composite system.

Operationally, the key question is whether a given interaction renders a relational sector dynamically accessible, so that changes in joint structure acquire physically traceable consequences. In this framework, a delocalized information flow is identified not by a taxonomic label, but by the redistribution of accessible relational support across subsystem partitions.

\subsection{Spatiotemporal extension}

The single-time framework above is sufficient for many of the controlled settings emphasized in this work. In history-dependent settings, however, the relevant hidden structure may be distributed not only across subsystems but also across time, so that an instantaneous state description becomes insufficient even at the level of the joint system. In this broader setting, the relevant organization is carried by temporally ordered transitions, interventions, or observations rather than by an instantaneous composite state alone. Such multi-time relational structure may be characterized, depending on the level of description, by the pseudo-density-operator formalism~\cite{2015_SRep_VV, 2024_PRL_MileGu, 2024_PRR_NelyNg}, quantum combs~\cite{2009_PRA_Comb}, or process tensors~\cite{2018_PRL_ProcessTensor, 2018_PRA_ProcessTensor}, which represent correlations or operational constraints distributed across temporally ordered interventions rather than being fully captured by any single-time state.

In this generalized sense, a delocalized information flow should not be understood as the literal motion of an information-bearing substance. Rather, it is the redistribution of operationally accessible relational structure across identifiable partitions of a composite description, whether those partitions are primarily spatial (single-time) or spatiotemporal (multi-time).

This broader notion will become useful in later sections when comparing substrate classes in which the most natural informational signatures may arise either through single-time relational redistribution or through temporally structured transition histories. We do not attempt a full process-theoretic treatment here. Rather, the purpose of this extension is only to identify the minimal conceptual generalization required when operationally relevant hidden structure is carried by temporally ordered correlations or transition histories rather than by a single-time composite state alone.

\section{Operational and dynamical characterization}

\subsection{Thermodynamic reciprocity}

The thermocoherent effect~\cite{Pusuluk_2021_PRR} provides a controlled transport realization of the single-time notion introduced above. In its original formulation, the effect does not merely show that correlations can modify heat transport. Rather, it exhibits a reciprocity relation between two distinct currents: a heat current and a coherence current. The first is familiar and thermodynamic. The second is relational: it is carried by shared coherence and is not reducible to any subsystem-local state variable. In the linear-response regime, these two currents can be written in Onsager-like form,
\begin{equation}
\begin{aligned}
\mathcal{J}_h &= L_{hh}\,\Delta\beta - L_{hc}\,\beta\,\Delta C,
\\
-\mathcal{J}_c &= L_{ch}\,\Delta\beta - L_{cc}\,\beta\,\Delta C,
\end{aligned}
\end{equation}
where \(\mathcal{J}_h\) and \(\mathcal{J}_c\) denote the heat and coherence currents, \(\Delta \beta\) and \(\beta\,\Delta C\) play the roles of thermal and coherence affinities, and \(\beta = 1/(k_B T)\) is the inverse temperature. The minus sign in the second equation reflects the sign convention adopted for the coherence current. The off-diagonal coefficients encode the thermocoherent coupling, and reciprocity is expressed by \(L_{hc}=L_{ch}\). Formally, this resembles thermoelectricity, but the analogy remains incomplete because the second current is not a classical charge-like flow.

In the underlying collision-model picture presented in Ref.~\cite{Pusuluk_2021_PRR}, the relevant coherence does not simply enter one subsystem as a local resource. Instead, it is redistributed across changing bipartitions of the global state, appearing transiently in different shared sectors without becoming attributable to any single subsystem in isolation. The flow is therefore not merely a bookkeeping device; it is the evolution of accessible relational structure under the interaction.

In this setting, the coherence current is best understood as the original explicitly current-like realization of a more general redistribution of thermodynamically accessible hidden relational structure. More general settings need not remain coherence-based, nor admit a unique scalar current representation, but the thermocoherent effect still provides the clearest transport archetype of this reciprocal logic.

\subsection{Thermal steering}

A complementary operational role appears in correlation-enabled Mpemba scenarios~\cite{Pusuluk_qMpemba}. Here, the key phenomenon is not a reciprocal current pair, but a change in relaxation hierarchy. Two initial states may share identical local thermodynamic descriptors -- such as local temperatures or local thermal marginals -- yet still relax differently because their global relational structures differ. In particular, a correlated locally thermal state may relax slower than an uncorrelated hotter state under the same admissible dynamics. In such cases, the faster or slower route toward equilibrium is not determined solely by local populations or energies, but by relational structure that remains hidden at the level of subsystem-reduced descriptions.

Taken together, these two archetypes show that the same general state-level logic can become operationally relevant in distinct ways: either through transport-coupled redistribution of relational structure or through altered relaxation ordering under fixed local thermodynamic data.

\subsection{Dynamical diagnostics}

Before turning to candidate neural substrates, it is useful to indicate how ordinary transport processes and the accompanying redistribution of relational structure may be characterized in practice. Once a candidate substrate is represented by an effective open-system model, both ordinary transport processes and hidden relational structure can, in principle, be analyzed within a common dynamical language.

At the level of a substrate \(S\) with an effective Hamiltonian \(\hat{H}_S\), Markovian dynamics of the relevant degrees of freedom may be written schematically in the form of a Lindblad-type master equation~\cite{BreuerAndPetruccione-2002}:
\begin{equation} \label{eq:general_lindblad}
\frac{d}{dt} \hat{\rho}_S(t)
= -\frac{i}{\hbar}[\hat{H}_S + \hbar \hat{H}_{LS},\hat{\rho}_S(t)] + \mathcal{D}(\hat{\rho}_S(t)) ,
\end{equation}
where the commutator term encodes the coherent contribution to the reduced system dynamics, while the second term, called the dissipator, accounts for irreversible exchange, dephasing, relaxation, or effective environmental driving. This separation is physically meaningful only to the extent that the effective generator is anchored in a microscopic model of the substrate--environment interaction. In purely phenomenological master equations, the partition between Lamb-shift-like coherent corrections (\(\hat{H}_{LS}\)) and dissipative terms is generally not unique~\cite{2013_Entropy_15_02100_Kosloff}, and physically relevant coherent renormalizations are often neglected altogether. For biologically realistic settings, where one aims to interpret the actual mechanisms underlying transport and relaxation, this limitation should be kept explicitly in mind. Moreover, when memory effects become operationally relevant, this local-in-time description should be replaced by a more general effective non-Markovian treatment appropriate to the transport process under consideration.

Once such a model is specified, ordinary transport observables can be defined as usual. For example, in the absence of explicit driving, the heat current into a subsystem \(S\) may be written as~\cite{2013_Entropy_15_02100_Kosloff, HeatCurrent}
\begin{equation} \label{Eq::HeatCurrent}
\mathcal{J}_h(t) = \mathrm{Tr}\!\left[ \hat{H}_S \, \mathcal{D}(\hat{\rho}_S(t)) \right],
\end{equation}
which reduces to a genuine heat flux in energy-preserving exchange settings such as the original thermocoherent model. Analogous current-like observables can be defined from the continuity equations or flux operators appropriate to particle, ion, excitation, or effective chemical transport in the substrate under consideration.

The same density operator may simultaneously encode relational resources that remain partially hidden from subsystem-reduced descriptions. Their role can be diagnosed using information-theoretic tools~\cite{modi2010unified, modi2012classical, adesso2016measures, streltsov2017colloquium}, including mutual information, relative-entropy-based diagnostics, and other physically motivated relational or operational measures, as well as more recent notions such as basis-independent coherence~\cite{2015_PRA_CoherenceAndDiscord} and correlated coherence~\cite{2016_PRA_CorrelatedCoherenceAndDiscordAndEnt}; for a more detailed discussion of representative quantifiers and their possible operational interpretations, see also Ref.~\cite{Pusuluk_2025_Entropy}. In transport settings of the thermocoherent type, these quantities can sometimes be reorganized into an explicit current-like object. A useful starting point is then the entropy production rate, which may be written as~\cite{2019_npj, 2019_PRL_Esposito_TDofInfoFlow, 2019_OSID, 2019_PRL_Esposito_PiWithBathCoh}
\begin{equation}
\Pi(t) = -\frac{d}{dt} S\!\left[\hat{\rho}_S(t)\,\|\,\hat{\rho}_S(\infty)\right],
\label{eq:general_entropy_production}
\end{equation}
in the presence of a single bath. Here, \(S[\cdot \, \| \, \cdot]\) is the quantum relative entropy which measures the distinguishability of two input density matrices.

A suitable decomposition of the entropy production rate may reveal relational contributions that are dynamically coupled to ordinary transport currents. More generally, however, a delocalized information flow need not always be representable by a unique scalar current. In many settings, it is better treated as a redistribution of accessible relational support whose effects are diagnosed indirectly through the dynamics. This distinction will matter in the next section, where some candidate neural substrates admit only partial or indirect operational access to the relevant hidden structure, even when no unique current-level decomposition is yet available.

\section{Generalization to neural matter} \label{Sec::Substrates}

The framework developed above is deliberately substrate-independent and, in principle, is not restricted to neural matter alone. More generally, the operational motifs recurring in the following substrate-specific sections, including classical reconfiguration of relational accessibility, transient quantum-to-classical residue formation near a functional bottleneck, and altered relaxation or resetting order (see Fig.~\ref{fig:operational_motifs}), may apply more broadly to biological settings where distinct transport processes can couple and partially hidden relational structure can become dynamically accessible under realistic environmental conditions.

We now turn to these candidate microscopic substrate classes in neural matter, not as uniquely privileged loci, but as a particularly rich and experimentally consequential arena in which electrical, chemical, ionic, and thermal processes coexist in unusually dense, strongly coupled, and functionally significant forms. The aim is therefore not to identify a single microscopic origin of cognition, but to examine physically grounded loci at which transport-coupled or history-dependent relational resources may become operationally consequential.

\subsection{Proton delocalization in hydrogen-bond networks}

Hydrogen-bond networks constitute one of the most natural microscopic substrate classes in which partially delocalized relational structure may arise in biological matter~\cite{Hassanali2016, Pusuluk_2018_PRSA, Pusuluk_2019_PRSA, Hassanali2021}.\footnote{Related behavior was sometimes described in our earlier work as ``back-and-forth proton tunneling'' or proton relocation. In retrospect, the more accurate emphasis is on proton delocalization across hydrogen-bonded donor--acceptor configurations, whose physical role is better understood as lowering the energy of the joint hydrogen-bonded arrangement rather than as a sequence of localized transfer events.} Across proteins, enzyme active sites, interfacial hydration layers, and confined water near functional biomolecules, the relevant donor--acceptor geometries are continuously reshaped by local electrostatics, molecular crowding, conformational fluctuations, and thermal motion. Under suitable geometric conditions, especially when donor--acceptor separation, bond angle, and local polarization transiently favor strong hydrogen bonding, the proton coordinate need not remain well described by a strictly localized occupation picture. Instead, it may become partially delocalized across donor and acceptor sites, thereby creating a microscopic relational resource that is not exhausted by any single local occupation variable.

A controlled proof-of-principle for this perspective was developed in our earlier open-system study of proton dynamics in a hexameric water-ice ring~\cite{Pusuluk_2019_PRSA}. There, proton motion in a geometrically constrained hydrogen-bond network was represented in a discrete local-occupation basis and then extended to a microscopic open-system description in which environmental effects were introduced through O--H stretch vibrations and O--O separation fluctuations, yielding a physically grounded reduced dynamics rather than a purely phenomenological master equation. The model was deliberately kept minimal, with only two effective free parameters, and was calibrated so that its steady-state behavior reproduced the experimentally reported proton-ordering crossover in hexagonal ice near \(58.9\,\mathrm{K}\) and \(73.4\,\mathrm{K}\). In the illustrative regime, these parameters corresponded to a proton-transfer scale of order \(1\,\mathrm{meV}\) and an interaction scale of order \(40\,\mathrm{meV}\), and the resulting transition behavior was noticeably sensitive to their values. Importantly, the stationary correlations displayed a structured temperature dependence: the classical correlations associated with proton motion within an individual hydrogen bond remained maximal across the temperature range considered, whereas the quantum correlations generated within those same bonds remained comparatively small but increased with temperature after \(60\,\mathrm{K}\) and reached its strongest values around the glass-transition regime near \(110\,\mathrm{K}\). Viewed retrospectively, this provides an early substrate-level illustration of the broader thermocoherent logic emphasized here: under suitable constraints, thermal environmental coupling can help render hidden relational structure dynamically accessible rather than acting only as a source of degradation.

That water-ice model should not be read as a literal model of neural tissue. The point is more modest and more useful. Water ice offers a clean and highly constrained hydrogen-bond environment in which proton delocalization and the resulting correlations can be analyzed under a controlled microscopic treatment. In this respect, it functions as a proof-of-principle for a broader substrate class. Structured or confined water near functional biomolecules is not equivalent to crystalline ice, but it can realize comparable geometric restrictions on donor--acceptor configurations and proton-accessible hydrogen-bond motifs. What matters here is not the specific phase of water, but the conjunction of geometric restriction, directional hydrogen-bond organization, strong environmental modulation, and the sensitivity of proton motion to collective local structure. These are the kinds of conditions under which hidden relational structure may become dynamically accessible.

A second and complementary lesson emerged from our earlier induced-fit model of enzyme recognition and tautomerization~\cite{Pusuluk_2018_PRSA}. There, two intermolecular hydrogen bonds between a substrate and a prototypical enzyme were analyzed within the same general discrete local-occupation framework, but now embedded in a biologically motivated setting in which the binding site of the enzyme undergoes a substantial \emph{classical conformational reshaping}. More concretely, the model was built around a geometry in which direct transfer is strongly suppressed at large donor--acceptor bond angles. By contrast, a conformationally induced transition-state geometry with a smaller bond angle can activate the relevant orbital overlap and proton-transfer pathway~\cite{Pusuluk_2018_PRSA}. The model again involved only two effective free parameters, chosen at chemically reasonable scales inspired by nucleotide-like hydrogen-bonded motifs. In the illustrative regime, these corresponded to energetic scales of order \(1\text{--}2\,\mathrm{eV}\) for the larger conformational reshaping and of order \(0.5\text{--}1\,\mathrm{eV}\) for the destabilization needed to access the activated proton-transfer geometry. The central result was not merely that proton delocalization could generate quantum correlations across the hydrogen bonds, but that the classically driven reshaping of the binding geometry could redistribute those correlations across all four hydrogen-bonded sites at the transition state, thereby making them operationally consequential for a subsequent reaction-like step~\cite{Pusuluk_2018_PRSA} (Fig.~\ref{fig:relational_accessibility}).

Even more significantly, the biologically most favorable regime in that analysis was not the one in which the redistributed quantum correlations remained long-lived. Rather, the key advantage arose when the correlation-enabled induced-fit pathway remained operationally favorable even after the transient quantum correlations had been converted into more robust classical correlations, whose entropic contribution further lowered the free energy of the transition state. More generally, this perspective helps avoid a common but potentially misleading expectation in quantum-biology discussions: functional relevance need not require the long-time preservation of fragile quantum coherence. In some biologically relevant regimes, the opposite may hold. Rapid quantum-to-classical transduction may be more functionally consequential than prolonged coherence preservation (Fig.~\ref{fig:transient_to_residue}). What matters is whether a transient microscopic quantum process can seed a hidden relational structure whose coarse-grained classical remnant remains dynamically accessible to subsequent transport, binding, switching, or reaction events. The model’s quantitative behavior points in the same direction. In the uncatalyzed neutral-channel approximation, the equilibrium population of the neutral transition state remained negligibly small, of order \(10^{-30}\), at physiological temperature. In the correlation-enabled induced-fit scenario, that transition-state population remained similarly negligible, but the alternative pathway became thermodynamically more accessible because the relevant free-energy barrier was effectively lowered into an intermediate regime between the substrate and product configurations. In the maximally redistributed idealized limit, the equilibrium weight of the product-like tautomer could then increase from \(0\) to \(0.5\) in the chosen model~\cite{Pusuluk_2018_PRSA}.

Taken together, these two earlier studies motivate a broader conclusion. Hydrogen-bond networks should be viewed not as vague ``quantum biology'' motifs, but as a substrate class for which one can already exhibit, in controlled models, the generation, redistribution, and partial degradation of hidden relational structure under realistic environmental influence. In the language of the present framework, they are plausible microscopic loci where proton delocalization can seed a partially delocalized relational resource, where environmental decoherence can transform quantum correlations into classically correlated remnants without necessarily destroying operational relevance, and where classical conformational or electrostatic reshaping can make such hidden structure consequential for transport or relaxation dynamics. Accordingly, hydrogen-bond-rich regions of neural matter may host microscopically generated relational resources whose coarse-grained consequences can persist beyond the lifetime of the underlying coherent superpositions. In such cases, the quantum contribution may function primarily as a transient preparatory stage that seeds a more robust classical relational residue.

\subsection{Aromatic $\pi$-networks and microtubule tryptophan architectures}

Aromatic $\pi$-electron networks constitute a second plausible substrate class in which partially delocalized relational structure may arise in biological matter. More generally, electron delocalization in aromatic systems can itself be interpreted as a form of quantum superposition shared across nonorthogonal atomic orbitals, suggesting a broader conceptual link between aromaticity and relational quantum structure~\cite{Aromatiq_Pusuluk_2024}. In proteins, aromatic residues can form spatially organized chromophoric assemblies whose transition dipoles support collective optical interactions over nanometric distances. Among these, tryptophan-rich architectures in microtubules are particularly suggestive, both because tryptophan residues exhibit strong near-ultraviolet absorption around \(280\,\mathrm{nm}\) and comparatively large transition dipoles, and because ordered microtubule geometry can place multiple chromophores in arrangements that favor collective excitonic effects~\cite{Craddock2014, celardo2019existence, babcock2024ultraviolet}. The resulting relational structure is not most naturally framed in terms of localized charge transport or long-lived electronic conduction in the ordinary solid-state sense. Rather, under suitable excitation conditions, it is more naturally associated with coherent delocalization within a single-excitation manifold, where excitation amplitude is distributed across multiple aromatic sites and where the corresponding informational structure is encoded in inter-site phase relations and the correlations they support.

This possibility was first articulated at the conceptual level in our recent perspective article~\cite{Pusuluk_2025_Entropy}, and was developed more explicitly through our open-system study of ultraviolet excitation dynamics in microtubule tryptophan networks~\cite{Pusuluk_2026_Entropy}. There, we considered a network of eight tryptophan chromophores within a tubulin dimer, using site-specific positions and dipole orientations extracted from structural data and following the radiative dynamics from picosecond to nanosecond timescales. Rather than stopping at the effective non-Hermitian descriptions commonly used in the microtubule excitonics literature, we reformulated the same dipole-coupled setting within a master equation-based open-system treatment, so that excitation loss and the redistribution of relational structure could be followed consistently in time. The resulting dynamics showed that the operational significance of excitation delocalization is not exhausted by whether an excitation simply survives longer or decays faster. More importantly, the same network can transiently support distinct regimes of routing, retention, and release, depending on how the excitation is initially prepared and on how the ordered geometry shapes access to rapidly emitting versus weakly emitting collective sectors.

A first lesson from that analysis is that the same aromatic network can behave either as a rapid exporting channel or as a transient correlation buffer. When the system is initialized in a bright collective mode, excitation and its associated relational structure are released on the ns scale, with most internal structure washed out within the first few nanoseconds. By contrast, when it is initialized in a dark collective mode, population and the associated hidden relational organization can remain trapped for tens of nanoseconds within the same eight-site tryptophan architecture before appreciable leakage occurs~\cite{Pusuluk_2026_Entropy}. In the language of the present framework, the same microscopic substrate can therefore realize distinct operational roles: under some preparations it behaves more like a correlation-emitting channel, whereas under others it behaves more like a temporary reservoir of partially delocalized relational structure.

A second lesson is that coherent delocalization itself can dynamically bias the system toward more correlation-preserving sectors. In our analysis, a fully phase-coherent excitation initially shared across all eight tryptophan sites did not merely produce transient interference patterns; over a \(0\text{--}15\,\mathrm{ns}\) window it also generated sustained internal relational structure while progressively shifting weight toward weakly radiative components of the collective spectrum~\cite{Pusuluk_2026_Entropy}. By contrast, a fully incoherent mixed excitation over the same sites failed to generate comparable internal redistribution of relational structure and instead decayed predominantly through outward radiative loss. This distinction is conceptually important for the present work. It shows that the relevant hidden relational resource is not reducible to spatial extent alone. What matters is whether the initial preparation contains the phase structure needed to access dynamically protected sectors in which correlations can be retained, rerouted, or selectively released. In this sense, excitation delocalization in aromatic $\pi$-networks is operationally significant not merely because it spreads amplitude, but because it can seed a relational organization that later evolution does not treat democratically.

A third and biologically more suggestive lesson is that local excitation events can induce site-selective routing. In the same study, site-localized initial excitations -- a more plausible proxy for photon absorption or oxidative excitation of individual aromatic residues -- were found to produce strongly site-dependent decay and retention patterns~\cite{Pusuluk_2026_Entropy}. Some tryptophan sites within the eight-chromophore dimer were much more effectively connected to protected, weakly radiative sectors and therefore supported longer internal retention, whereas others coupled more efficiently to rapidly emitting channels. Embedding the same focal dimer into progressively larger structures, from a two-tubulin system to one-spiral and two-spiral segments, further reshaped which chromophore pairs became dynamically relevant and how strongly local initial conditions continued to matter. The broader implication is that ordered aromatic architectures need not merely support delocalization in an abstract sense; they can also implement a form of structurally constrained route selection, in which local geometry and initial site specificity jointly determine whether relational structure is preferentially retained, redistributed internally, or exported outward, consistent with the more general motif of reconfigured relational accessibility illustrated schematically in Fig.~\ref{fig:relational_accessibility}.

A fourth lesson concerns the role of scale and disorder. In larger ordered assemblies, both outward export and internal retention channels became more pronounced, indicating that increasing architectural scale can strengthen the separation between rapidly emitting and weakly emitting collective sectors~\cite{Pusuluk_2026_Entropy}. In the ideal ordered construction, one circumferential spiral contains 13 tubulin dimers and spans approximately \(22.4\,\mathrm{nm}\) in diameter, while the largest assemblies we examined extended to 100 such spirals. Across this size range, bright collective modes became substantially faster and dark modes substantially longer-lived, with the latter extending into the millisecond range in the ordered limit. At the same time, static energetic disorder and structural disorder reduced long-range redistribution and suppressed the overall transfer of hidden relational structure, consistent with the intuitive expectation that environmental variability degrades coherent organization~\cite{Pusuluk_2026_Entropy}. Yet the effect was not simply all-or-nothing. Disorder diminished long-range flow without erasing all protected retention pathways. This is important for the present framework because it reinforces a recurring theme already seen in other substrate classes: realistic biological noise need not eliminate operationally relevant hidden structure altogether. Rather, it can reshape which sectors remain accessible, how long they persist, and over what spatial or temporal range their consequences remain detectable.

Finally, the same tryptophan-network analysis also suggested that the relevant structure may not be exhausted by purely memoryless state-level descriptions. In addition to equal-time redistribution within a focal dimer, embedding that dimer in a larger spiral generated short-time revivals consistent with information returning from the surrounding tubulin environment to a local subsystem~\cite{Pusuluk_2026_Entropy}. While we do not claim that such behavior by itself establishes a full process-level description, it does indicate that short-time retention and release of relational structure may already depend on more than instantaneous local populations. In that limited but useful sense, aromatic $\pi$-networks provide not only a candidate substrate for single-time excitation delocalization, but also a plausible bridge toward the broader process-level perspective developed earlier in this work, where the operational relevance of hidden structure may extend across ordered times as well as across spatial partitions.

Taken together, these observations support a cautious but concrete conclusion. Aromatic $\pi$-electron architectures in neural matter -- particularly organized tryptophan networks in microtubules -- should not be treated as evidence for robust macroscopic quantum cognition, nor as a unique microscopic origin of neural information processing. Rather, they provide one physically structured substrate class in which one can already exhibit, under explicit open-system dynamics, the generation, routing, buffering, and selective release of hidden relational structure within realistic geometric and dissipative constraints. This broader interpretation is also consistent with related work on biomolecular photoprocesses, where energetic quantum coherence has been shown to enhance photoisomerization efficiency in multi-molecule settings~\cite{burkhard2024boosting}. In the language of the present framework, they are plausible microscopic loci where excitation delocalization can seed a partially delocalized relational resource, and where ordered geometry can bias that resource toward retention or outward export. Local site specificity can then implement route selection, while environmental disorder can limit but need not completely erase operational relevance. Their significance therefore lies not in proving long-lived quantum computation in the brain, but in showing that biologically embedded aromatic networks can transiently organize relational resources in ways that may influence downstream transport, local interaction landscapes, or mesoscale coordination.

\subsection{Phosphate-rich motifs and tetrahedral spin buffering}

Phosphate-rich molecular motifs provide a distinct and conceptually complementary substrate class in which hidden relational structure may be protected, redistributed, or rendered dynamically accessible through geometry-dependent buffering rather than through particle delocalization. In biological environments, phosphate groups are ubiquitous in ATP-related chemistry, phospholipid membranes, phosphorylation cycles, and calcium-phosphate assemblies. Within speculative quantum-cognition proposals, they have also attracted particular attention because phosphorus nuclei carry spin-$\tfrac{1}{2}$ and are therefore, in principle, less susceptible to certain electric-field-induced decoherence channels than higher-spin nuclei~\cite{fisher2015quantum,weingarten2016new}. The present framework does not require adopting the stronger claims of the Posner model. The more modest and physically relevant point is that phosphate-rich motifs naturally realize local interaction geometries in which internal degrees of freedom can be partially shielded from environmental degradation by the surrounding network architecture.

A controlled proof-of-principle for this idea was developed in our earlier study of open spin clusters with geometry-dependent buffering~\cite{Pusuluk_2024_PRA_Posner1}. There, we considered a central spin coupled to a surrounding network of buffer spins, with the buffer spins exposed to local thermal environments while the central spin remained isolated from direct dissipation. Although the model was intentionally simplified and did not explicitly model a full phosphate or Posner molecule, it was designed to isolate a precise question: how does interaction geometry affect the persistence of coherence in an embedded microscopic degree of freedom? The answer was nontrivial. Greater protection did not arise simply by increasing the number of surrounding spins. Instead, the decisive factor was the connectivity and geometry of the buffer network. For a fixed number of buffer spins, maximal planar connectivity yielded the longest coherence-preservation times, but the overall optimum was not obtained by the largest cluster considered. Rather, a four-spin buffer network with maximal connectivity -- corresponding to a tetrahedral arrangement -- provided the most effective protection against environmental degradation~\cite{Pusuluk_2024_PRA_Posner1}.

This result is particularly significant in the present context because it already points beyond a purely local notion of coherence storage. The protected quantity in that model was a central-spin coherence, but its persistence was controlled by a surrounding relational shell whose internal coupling structure redistributed environmental influence before it could erase the central resource. In other words, the relevant physical lesson was not simply that ``coherence survives longer'', but that a specific geometry can delay the thermodynamic accessibility of coherence-erasing pathways. In the original analysis, this was reflected not only in coherence measures but also in the delayed exchange of the heat required to erase the central coherence, suggesting that the buffering geometry reorganizes the route by which dissipation becomes operationally effective~\cite{Pusuluk_2024_PRA_Posner1}. For example, in the pure-dephasing comparison reported there, the tetrahedral four-spin buffer roughly doubled the long-time survival scale of the central-spin coherence relative to the minimally connected alternative for the same cluster size, indicating that the advantage is quantitatively substantial and genuinely geometric rather than merely qualitative. Viewed from the standpoint of the present framework, this is naturally interpreted as a geometry-controlled persistence of hidden relational structure.

A second and more directly relational lesson emerged in our later perspective study, where the same toy-model logic was extended from single-cluster coherence preservation to a two-cluster scenario with entangled central spins~\cite{Pusuluk_2025_Entropy}. There, the goal was not to claim a realistic microscopic model of Posner chemistry, but to test whether the same buffering architecture that protects a local coherent resource can also help preserve an explicitly distributed one. The answer was again suggestive: the tetrahedral arrangement remained the most favorable geometry, now not only for preserving coherence in a single protected degree of freedom, but also for prolonging entanglement between central spins embedded in separate clusters~\cite{Pusuluk_2025_Entropy}. This extension matters conceptually because it shifts the emphasis from local resource retention to the persistence of relational structure across distinct units. In the language of the present work, the same substrate geometry can support both the buffering of single-site coherence and the longer-lived accessibility of inter-unit hidden structure, consistent with the more general operational motif of transiently quantum structure leaving a more persistent relational residue, illustrated schematically in Fig.~\ref{fig:transient_to_residue}.

The theoretical interpretation proposed in that later analysis is also aligned with the present framework. The advantage of the tetrahedral arrangement was not treated as a mere numerical coincidence, but as a consequence of how the geometry reorganizes the effective spectrum and partially isolates the central degrees of freedom from the most destructive dissipative channels~\cite{Pusuluk_2025_Entropy}. This is exactly the kind of substrate-level mechanism that the present work seeks to identify more generally: not a claim that cognition depends on a specific exotic molecule, but the more modest proposal that certain local architectures can act as relational buffers, allowing microscopically encoded structure to remain dynamically consequential for longer than naïve local-noise arguments would suggest.

Taken together, these two studies motivate a broader conclusion. Phosphate-rich motifs should not be invoked merely as speculative ``quantum cognition'' symbols, nor should the present argument be read as an endorsement of the full Posner program. The stronger and more defensible lesson is that tetrahedral or near-tetrahedral local interaction geometries -- of the sort naturally associated with phosphate-centered coordination environments -- can, in controlled models, act as protective relational shells. Such architectures can buffer local coherent resources, delay thermodynamically effective erasure, and in some cases prolong distributed correlations between separated protected degrees of freedom. In the language of the present framework, phosphate-rich regions of neural matter are therefore plausible microscopic loci where geometry-sensitive hidden relational structure may persist longer than the underlying fragile quantum signatures that first seed it. This does not imply long-lived macroscopic quantum cognition. It implies only that local molecular geometry may help determine which microscopic relational resources remain available for subsequent transport, binding, switching, or coordination events.

\subsection{Ion channels as barrier-structured transduction interfaces}
\begin{figure*}[t]
    \centering

    % Top row
    \begin{subfigure}[t]{0.52\textwidth}
        \centering
        \includegraphics[width=\linewidth]{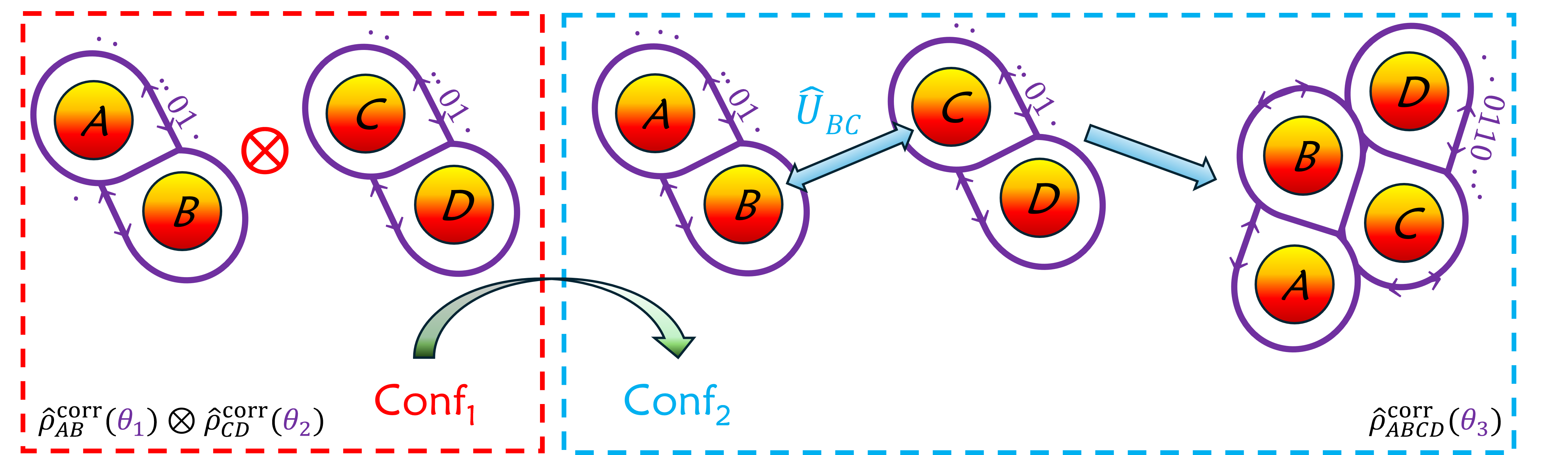}
        \caption{\textit{Classical reconfiguration of relational accessibility}}
        \label{fig:relational_accessibility}
    \end{subfigure}

    \vspace{0.8em}

    % Bottom row
    \begin{subfigure}[t]{0.49\textwidth}
        \centering
        \includegraphics[width=\linewidth]{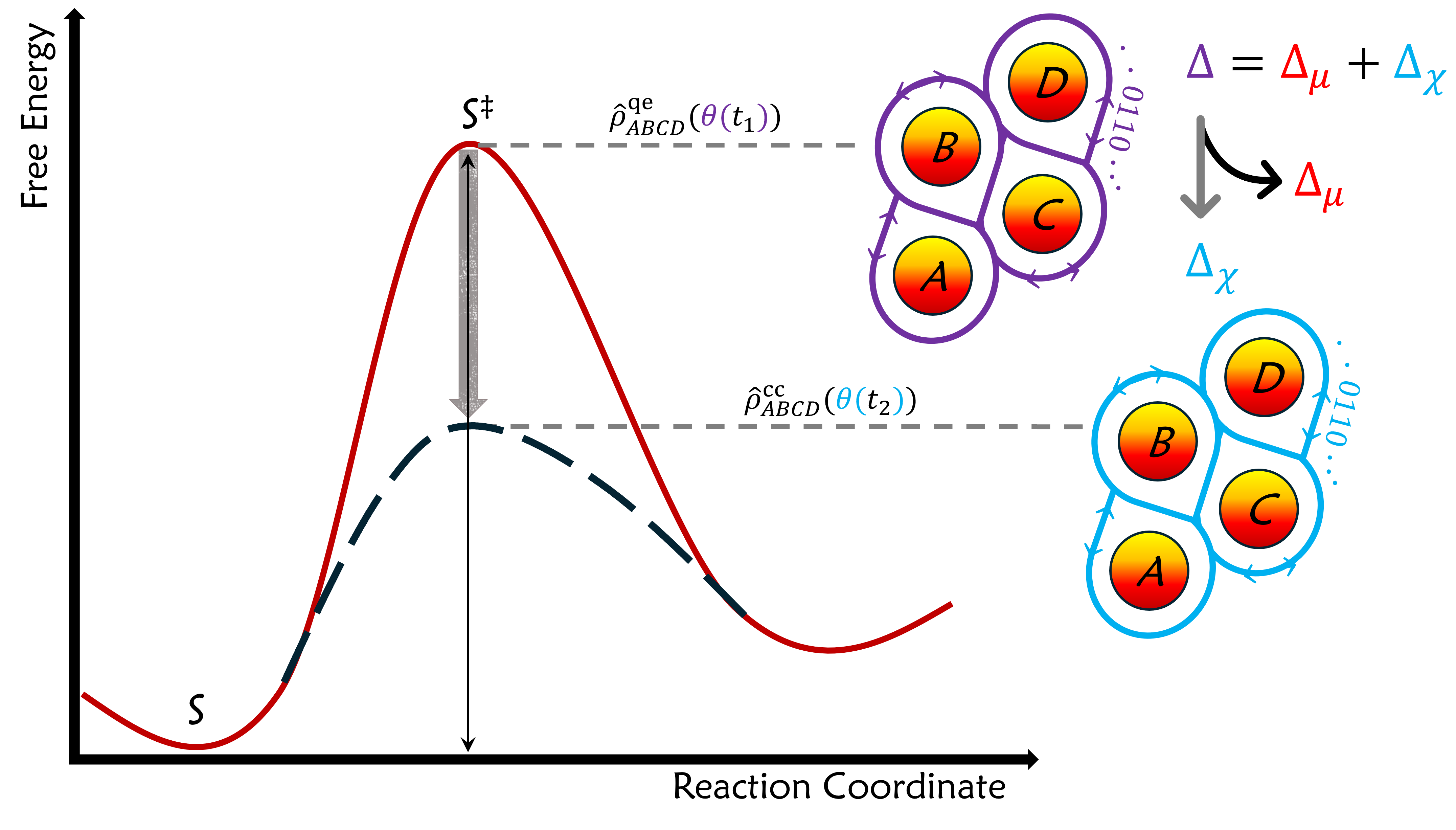}
        \caption{\textit{From transient entanglement to classical relational residue}}
        \label{fig:transient_to_residue}
    \end{subfigure}
    \hfill
    \begin{subfigure}[t]{0.48\textwidth}
        \centering
        \includegraphics[width=\linewidth]{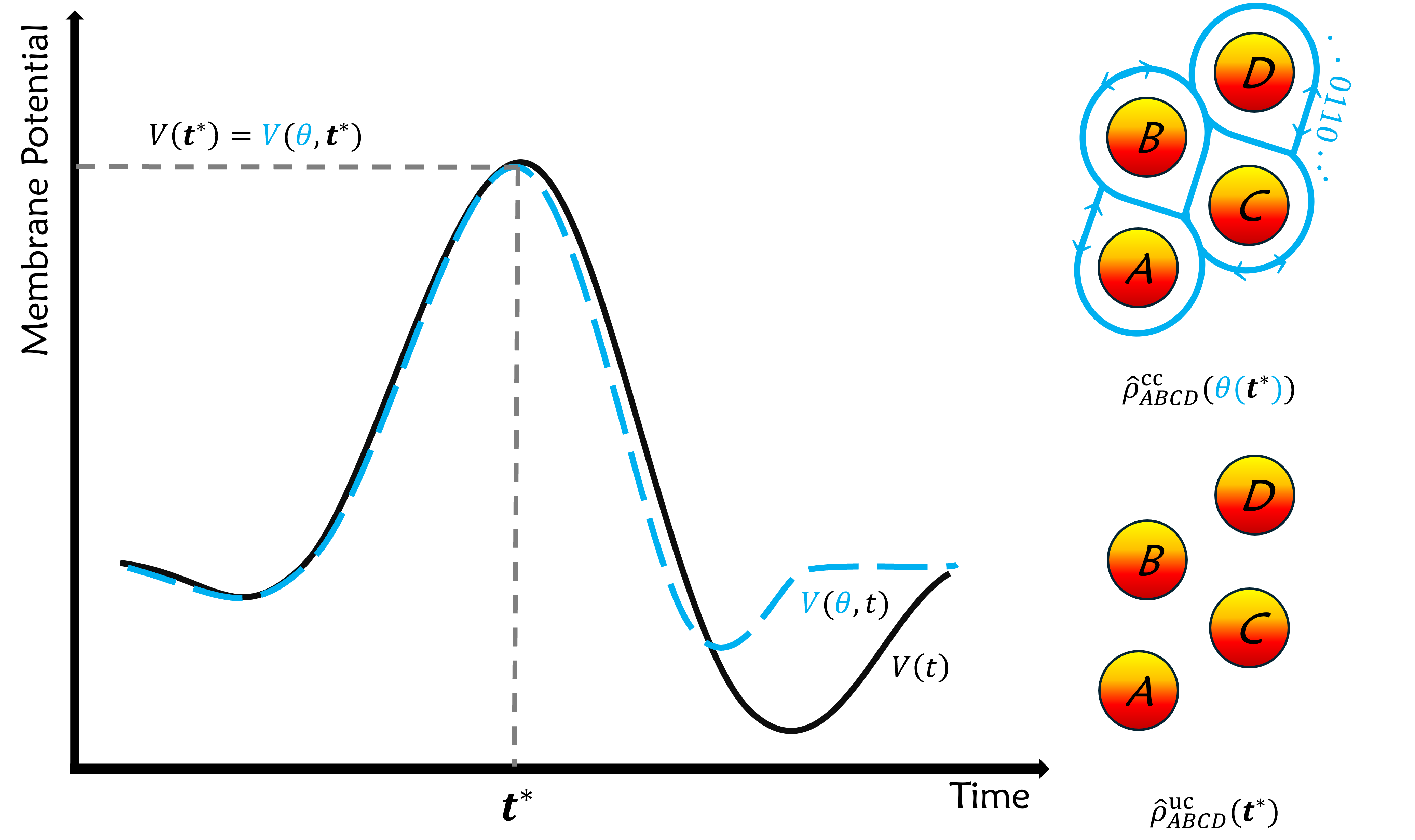}
        \caption{\textit{Mpemba-type membrane resetting via hidden relational structure}}
        \label{fig:mpemba_resetting}
    \end{subfigure}

    \caption{\justifying
    Schematic summary of three substrate-independent operational motifs by which hidden relational structure may become dynamically consequential in biological or neural matter. 
    (a) A classically driven conformational or geometric reconfiguration (Conf$_1 \to$ Conf$_2$) can alter which relational sectors of a composite system become dynamically accessible under the effective interaction geometry, allowing hidden relational support to be redistributed across new subsystem partitions. 
    (b) A short-lived quantum contribution associated with the entanglement-linked sector $\Delta_\mu$ need not remain long-lived to be functionally relevant: near a transition region or reaction bottleneck, transiently accessible quantum structure can be converted into a more robust classically correlated residue $\Delta_\chi$, leaving a persistent relational bias after the underlying quantum correlations have decohered. 
    (c) Two preparations with comparable coarse-grained local membrane descriptors can nonetheless exhibit distinct membrane-potential recovery trajectories if they differ in hidden relational structure; schematically, a relational contribution $\theta$ that becomes dynamically consequential near the resetting transition can alter the ordering or timing of repolarization, in analogy with correlation-enabled Mpemba-type relaxation. This panel is intended as a coarse-grained operational motif rather than a literal electrophysiological model. 
    Together, the three panels highlight a central theme of the present framework: biologically relevant hidden relational resources need not rely on long-lived coherence alone, but may also emerge through classical reconfiguration, transient quantum-to-classical transduction, and altered relaxation or resetting order at the coarse-grained level.
    }
    \label{fig:operational_motifs}
\end{figure*}

The preceding substrate classes are best viewed as candidate microscopic media in which partially delocalized relational structure may arise, persist, or be redistributed. Ion channels occupy a different and, in neurophysiological terms, arguably more privileged role. They provide one of the clearest interfaces through which such hidden structure could become functionally consequential for membrane excitability, gating, spike initiation, and signal transmission. In this sense, ion channels need not be the cleanest setting in which to establish a fully characterized delocalized information current in the strict thermocoherent sense; rather, they are among the most plausible sites at which microscopically generated relational resources can be transduced into physiologically relevant transport outcomes.

This privileged role follows from the physical structure of ion-channel conduction itself. Ions do not simply drift through homogeneous conduits. Rather, they traverse highly constrained pathways in which steric confinement, hydration structure, local electrostatics, and protein conformational fluctuations generate a sequence of metastable occupancy configurations separated by dynamically modulated free-energy barriers. Potassium channels are especially instructive in this regard: they combine near diffusion-limited throughput with extraordinary selectivity, forcing transport to occur under unusually tight geometric and energetic constraints. Recent analyses of the KcsA selectivity filter support the view that the relevant conduction landscape is not static but is continuously reshaped by neighboring ions, confined water molecules, and the fluctuating protein environment. In such a dynamically reconfigured multi-ion setting, a purely classical over-the-barrier description may be incomplete in some regimes, and a tunneling contribution may need to be considered~\cite{Pusuluk_2026_PBMB}. In the KcsA proof-of-principle analysis, the thermally averaged time scale remained within a biologically relevant regime: at \(300\,\mathrm{K}\), a single K$^+$ ion was estimated to spend about \(5\,\mathrm{ns}\) in the selectivity filter due to the barrier structure, reinforcing that the proposed hidden-structure effects need not be dismissed as parametrically disconnected from channel function~\cite{Pusuluk_2026_PBMB}. In the present framework, the soft-knock versus hard-knock distinction may therefore be read more abstractly as a difference in which multi-site occupancy sectors become dynamically accessible and in which forms of inter-site or inter-event relational structure may become operationally relevant.

In this respect, ion channels may support a form of history-dependent relational steering even when a strict current-level decomposition is not yet available. The relevant hidden structure need not reside primarily in a persistent equal-time spatial correlation between simultaneously occupied sites. More plausibly, it may be distributed across ordered conduction histories: barrier crossing, transient trapping, partial reflection, and occupancy-dependent route selection can render later transport events sensitive to the detailed causal structure of earlier ones. A possible tunneling contribution, if present in a given regime, may then act less as a source of long-lived spatially delocalized coherence and more as a microscopic modifier of future route accessibility. In this perspective, a tunneling transition can be read heuristically not merely as a spatial superposition between two sites, but as a coherent superposition of alternative transport histories. In a minimal heuristic notation, one may write
\begin{equation}
|\Psi_{\mathrm{hist}}\rangle
=
r\,|A_{t_1}\!\to\!A_{t_2}\rangle
+
\tau\,|A_{t_1}\!\to\!B_{t_2}\rangle ,
\label{eq:history_superposition_compact}
\end{equation}
where the first branch corresponds to remaining in the initial well and the second to barrier traversal toward a downstream site. An explicit occupation-number embedding of the same idea in a temporally factorized tensor-product description is
\begin{equation}
\label{eq:history_superposition_explicit}
\begin{aligned}
|\Psi_{\mathrm{hist}}\rangle
=
&r\,|1\rangle_{A,t_1}|0\rangle_{B,t_1}|1\rangle_{A,t_2}|0\rangle_{B,t_2}\\
&\quad+
\tau\,|1\rangle_{A,t_1}|0\rangle_{B,t_1}|0\rangle_{A,t_2}|1\rangle_{B,t_2}.
\end{aligned}
\end{equation}
This representation should be understood as heuristic rather than foundational: its purpose is not to redefine tunneling, but to make explicit that a tunneling process may be interpreted as a coherent superposition of alternative transport histories. In that sense, the relevant hidden structure is relational not only across sites but also across ordered times. An operationally complete treatment of such history dependence is more naturally given by pseudo-density-operator~\cite{2015_SRep_VV, 2024_PRL_MileGu, 2024_PRR_NelyNg}, quantum comb~\cite{2009_PRA_Comb}, or process-tensor~\cite{2018_PRL_ProcessTensor, 2018_PRA_ProcessTensor} descriptions, in which intermediate interventions can reveal whether later transport statistics depend only on instantaneous local populations or on a genuinely multi-time relational structure.

Accordingly, ion channels suggest a biologically central setting in which hidden relational structure may become operationally relevant not only through single-time redistribution across occupancy sectors, but also through multi-time organization encoded in ordered conduction histories. In this multi-time setting, the relevant resource may be invisible to local one-time descriptors such as single-site populations or instantaneous electrochemical gradients, yet still bias waiting-time statistics, route selection, effective barrier accessibility, or relaxation ordering, in close analogy with the coarse-grained membrane-resetting motif illustrated schematically in Fig.~\ref{fig:mpemba_resetting}. Because ion-channel conduction directly shapes transmembrane currents and local charge redistribution, it also provides a natural bridge to mesoscale electromagnetic organization. In this sense, field-level structure may function not as the primary microscopic carrier of the relevant relational resource, but as a coarse-grained readout and possible feedback layer through which channel-level history dependence and transport selectivity can influence larger-scale neural coordination.

Unlike some of the other candidate substrate classes discussed above, the present ion-channel example does not yet come with a complete open-system decomposition of physical and relational currents. Its role here is therefore deliberately different: it identifies a barrier-structured transduction interface in which hidden relational structure can plausibly become neurophysiologically consequential, while also motivating future analyses based on explicit open-system models, multi-time correlation measures, pseudo-density-operator constructions, or process-tensor descriptions.

\section{Multiscale thermocoherent organization in neural matter}

\begin{figure*}[t]
    \centering
    \includegraphics[width=.68\textwidth]{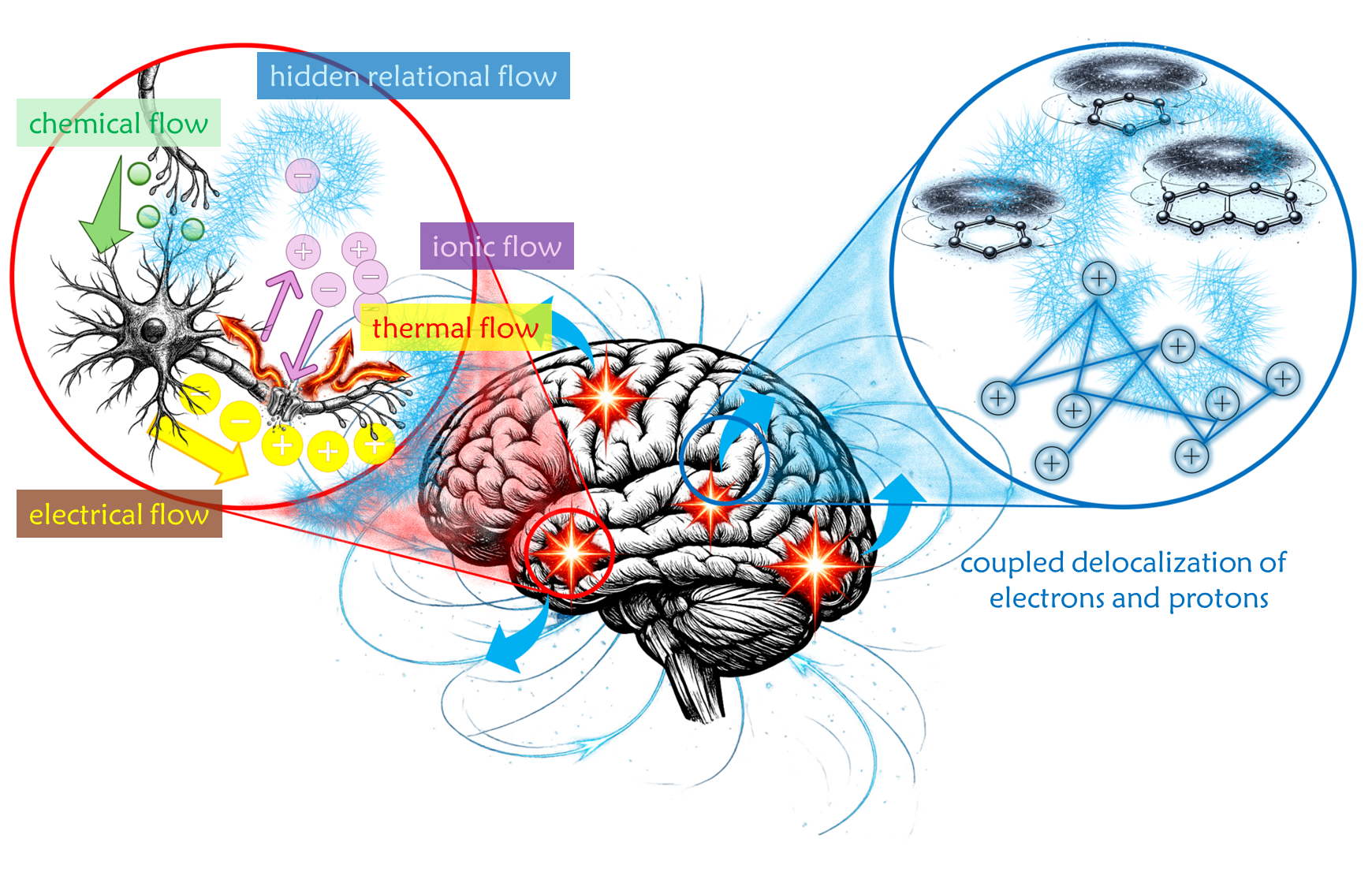}
    \caption{\justifying Schematic illustration of coupled transport processes in neural matter and their possible multiscale reorganization. Electrical, chemical, ionic, thermal, and correlation flows may generate, transduce, or redistribute hidden relational structure across suitable microscopic substrates. Such substrate-level contributions -- including coupled proton/electron delocalization, aromatic \(\pi\)-architectures, and geometry-dependent buffering motifs -- may then coarse-grain into larger-scale thermocoherent organization across neural tissue.}
    \label{Fig::ThermoCohBrain}
\end{figure*}

Taken together, the substrate classes discussed above suggest that heat flow is not unique in its capacity to participate in the generation, transduction, buffering, or redistribution of hidden relational structure. In neural matter, electrical, chemical, ionic, and thermal transport processes may each, under suitable microscopic conditions, support such structure, as illustrated schematically in Fig.~\ref{Fig::ThermoCohBrain}. The relevant structures need not all take the same strict form. In some cases, they arise through transient microscopic delocalization and later survive as more robust classical relational remnants; in others, they are buffered by local geometry, routed through ordered excitonic architectures, or rendered operationally consequential at barrier-structured interfaces. What unifies these possibilities is not a single microscopic mechanism, but the more general fact that partially hidden relational resources can become dynamically accessible under the appropriate interaction geometry, spectral structure, or temporal ordering.

Several common lessons emerge from the candidate substrates examined above. First, environmental influence need not act only as a source of degradation. In hydrogen-bonded settings, thermal coupling can in some regimes help render hidden relational structure dynamically accessible rather than simply suppressing it. Second, classical reshaping can itself become a transduction mechanism: in hydrogen-bond networks, classically driven conformational changes can redistribute initially quantum correlations and convert them into more robust classical relational residues that remain operationally useful. Third, ordered aromatic $\pi$-architectures show that the same microscopic network can behave either as a rapid exporting channel or as a transient correlation buffer, depending on preparation, geometry, and radiative sector accessibility. Fourth, phosphate-rich motifs indicate that local interaction geometry can act as a protective shell, delaying thermodynamically effective erasure and prolonging the accessibility of fragile local or distributed relational resources. Finally, ion channels suggest that some of the most biologically central consequences of hidden structure may arise not where that structure is cleanly generated, but where it is transduced into physiologically relevant transport outcomes through barrier-structured, history-dependent dynamics.

This viewpoint suggests a multiscale bridge between microscopic physics and mesoscale neural organization. At small scales, hidden relational resources may reshape local energy landscapes, transport selectivity, route accessibility, and relaxation orderings. At larger scales, their effects may coarse-grain into altered effective transition structure, metastable stabilization, modified waiting-time statistics, or changes in the routing and persistence of neural activity. In this respect, the candidate substrate classes considered above play complementary rather than competing roles: hydrogen-bonded proton networks can seed and transduce transient quantum-to-classical relational resources; aromatic $\pi$-electron architectures can route, buffer, and selectively release excitation-borne correlations; phosphate-rich motifs can geometrically protect vulnerable microscopic resources; and ion channels can act as privileged transduction interfaces where multi-time relational structure becomes functionally consequential for signaling.

At the same time, this substrate-centered perspective remains compatible with, but not reducible to, field-based~\cite{Pockett2000, McFadden2000, mcfadden2002synchronous, mcfadden2002b, Pockett2012, mcfadden2013a, mcfadden2013b, McFadden2020} and systems-level~\cite{Fries2005, Fries2015} proposals of cognition, including electromagnetic field-centered approaches. Large-scale electromagnetic organization may indeed correlate with, or help stabilize, certain cognitive states, yet we do not treat such fields as the primary microscopic resource. Instead, they are more naturally interpreted as emergent, coarse-grained readouts or feedback layers of deeper nonequilibrium organization shaped by coupled electrical, chemical, ionic, and thermal transport processes together with the hidden relational structure they generate and reorganize. In this sense, field-level order may be understood not as the sole microscopic origin of cognitive organization, but as one mesoscale manifestation of a deeper multiscale thermocoherent architecture.

The central claim of the present work is therefore narrower than a full theory of consciousness but more concrete than a purely metaphorical appeal to ``quantum effects in the brain''. We propose that microscopically generated hidden relational resources, including both quantum and classical forms, may influence thermodynamically constrained transport, relaxation, and signaling processes in neural matter. Their significance need not lie in sustaining fragile quantum signatures at macroscopic scales, but in shaping the local and mesoscopic dynamics out of which cognitive organization is built. The aim of this framework is not to replace existing neuroscientific models, but to supply a physically grounded and experimentally constrained language in which resource-theoretic ideas from quantum information thermodynamics can be connected to multiscale information flow in the brain. 

More broadly, although neural matter provides the central focus of the present perspective, the framework developed here is not intrinsically restricted to the brain. It may also apply to other biological systems in which transport, structured environmental interactions, and partially hidden relational organization jointly shape relaxation, route selection, or multiscale coordination.

\section{Toward falsifiable consequences}

A framework of the kind proposed here is scientifically useful only if it yields empirically discriminating consequences. We therefore emphasize that the present work should not be read as a generic invitation to search for ``quantum effects in the brain'' in an unconstrained sense. The more specific claim advanced here is that, if hidden relational structure becomes thermodynamically accessible in neural matter, its most plausible signatures will appear not primarily as directly measurable long-lived microscopic quantum states, but as structured and reproducible deviations in coarse-grained transport, relaxation, timing, and coordination observables. The relevant question is therefore not simply whether one can detect microscopic coherence, but whether one can identify mesoscopic signatures that are more naturally explained by hidden accessibility constraints than by purely local, memoryless effective descriptions.

A particularly concrete arena in which the present framework could generate experimentally discriminating predictions is post-activation recovery, including the reset phase that follows strong local depolarization. If ion-channel configurations, nearby proton-coupled microenvironments, or related microscopic substrate degrees of freedom can transiently encode hidden relational structure without measurably changing the usual coarse local descriptors, then two post-activation states that appear similarly perturbed at the level of membrane voltage, local ionic imbalance, or effective thermal load need not relax back toward functional readiness at the same rate. In that case, one may obtain a \emph{Mpemba-like reset asymmetry} (Fig~\ref{fig:mpemba_resetting}): a state that is not obviously ``closer'' to recovery according to standard local observables may nonetheless return faster because hidden relational structure alters the dynamically accessible relaxation route. At present, this should be regarded only as a falsifiable scenario rather than an established mechanism. Its value lies in motivating substrate-specific open-system models and controlled perturbation protocols that compare recovery-time predictions under matched coarse local observables but systematically varied hidden relational preparations.

More generally, the strongest tests of the present framework are unlikely to rely on a single spectacular marker. Rather, they should seek correlated families of anomalies: structured perturbation-dependent changes in recovery timing, route selection, relaxation ordering, waiting-time statistics, metastable persistence, and, at larger scales, field-level coordination patterns. The evidential burden should therefore not rest on the direct observation of a fragile microscopic quantum signature, but on whether multiple observables shift together in the direction expected if hidden relational structure is modifying thermodynamic accessibility. This is precisely what makes the present proposal falsifiable: it predicts specific patterns of dependency across scales, not merely isolated deviations.

Before any decisive experimental program can be justified, however, the more immediate need is sharper theory. The candidate substrate classes discussed here are not yet on equal mechanistic footing, and the present work does not claim to provide a complete open-system account for all of them. The next step is therefore to construct more explicit substrate-specific dynamical models, identify discriminating observables under controlled perturbations, and determine which signatures genuinely exceed what can already be captured by carefully calibrated local effective descriptions.

\section{Conclusion}

In this work, we have proposed a thermocoherent perspective on the physical basis of information flow in neural matter. Starting from the thermocoherent effect and recent correlation-enabled relaxational phenomena, we argued that the operational relevance of relational structure depends less on whether it is labeled as entanglement, discord, or classical correlation, and more on whether it becomes dynamically accessible under the geometry and nonequilibrium constraints of the underlying physical process. From this standpoint, information flow need not be treated as an abstract coding primitive detached from matter, nor as something exhausted by subsystem-local state variables. It may instead be grounded in partially hidden relational structure that can be generated, redistributed, transformed, and occasionally amplified by irreversible transport.

This perspective naturally motivates a multiscale picture of neural organization. Electrical, chemical, ionic, and thermal processes need not be regarded as merely parallel carriers of separate forms of neural activity. Under suitable microscopic conditions, each may generate or transduce partially delocalized relational resources whose coupling can reshape local accessibility, route selection, relaxation ordering, and metastable persistence, with possible coarse-grained consequences for mesoscale coordination. In this sense, candidate substrate classes such as hydrogen-bonded proton networks, aromatic $\pi$-electron architectures, phosphate-rich motifs, and ion-channel interfaces are not invoked as speculative replacements for established neuroscience, but as physically grounded loci where distinct operational roles of hidden relational structure may become biologically consequential: hydrogen-bonded networks can seed transiently accessible relational resources and quantum-to-classical residues, aromatic architectures can route or buffer excitation-linked structure, phosphate-rich motifs can provide geometry-dependent buffering of fragile microscopic resources, and ion-channel interfaces can transduce such resources into physiologically relevant, potentially history-dependent transport outcomes.

We have also emphasized what the present framework does \emph{not} claim. It is not a theory of consciousness, not a claim that long-lived macroscopic quantum coherence is generically present in the brain, and not a dismissal of successful systems-level or field-based descriptions. Rather, it is a proposal that some mesoscale regularities relevant to cognitive dynamics may ultimately be better understood when viewed as coarse-grained manifestations of deeper transport-coupled relational organization. In that sense, field-level structure may be interpreted not as the unique microscopic seat of cognition, but as an emergent coordination layer or readout of more deeply distributed thermodynamic constraints.

The value of the present work therefore lies less in establishing any single substrate mechanism than in formulating a physically constrained and falsifiable research program. If the perspective developed here is correct, its earliest support will likely come not from spectacular direct observations of fragile microscopic quantum states, but from reproducible, structured deviations from overly local effective descriptions across transport, timing, relaxation, metastability, and cross-scale coordination. If such signatures fail to materialize under increasingly refined theoretical and experimental scrutiny, then the framework should be correspondingly revised or abandoned. If they do emerge, however, they may help connect microscopic physics to cognitive dynamics in a way that is both more conservative and more physically explicit than either unconstrained quantum-cognition proposals or purely abstract information-processing narratives.

\begin{acknowledgments}
The author thanks Vlatko Vedral, \"{O}zg\"{u}r E. M\"{u}stecapl{\i}o\u{g}lu, Lea Gassab, and Eda Alemdar for helpful comments on earlier versions of this manuscript and for encouraging discussions.
\end{acknowledgments}

%\bibliography{References}


\end{document}